\documentclass[aps,prx,twocolumn,superscriptaddress]{revtex4-1}

\usepackage{blindtext}
\usepackage{amsmath,amssymb,mathrsfs}
\usepackage{natbib}
\usepackage{subfigure}
\usepackage{tabularx}
\usepackage{epsfig}
\usepackage{longtable}
\usepackage{amsfonts}
\usepackage{rotating}
\usepackage{subfigure}
\usepackage{amsmath}
\usepackage{comment}
\usepackage{bbold} 
\usepackage{color}
\usepackage{braket}

\usepackage[unicode=true,bookmarks=true,bookmarksnumbered=false,bookmarksopen=false,breaklinks=false,pdfborder={0 0 1},backref=false,colorlinks=true]{hyperref}

\hypersetup{linkcolor=magenta,urlcolor=blue,citecolor=blue,pdfstartview={FitH},hyperfootnotes=false,unicode=true}

\def\be{\begin{equation}}
\def\ee{\end{equation}}
\def\bea{\begin{eqnarray}}
\def\eea{\end{eqnarray}}
\def\nn{\nonumber}

\def\be{\begin{equation}}
\def\ee{\end{equation}}
\def\bea{\begin{eqnarray}}
\def\eea{\end{eqnarray}}
\def\nn{\nonumber}

\newcommand{\bDiamond}{\mathbin{\Diamond}} 

\begin{document}

\title{Kernel-Function Based Quantum Algorithms for Finite Temperature Quantum Simulation}

\author{Hai Wang$^{\dagger}$} 
\affiliation{State Key Laboratory of Surface Physics, Institute of Nanoelectronics and Quantum Computing, and Department of Physics, Fudan University, Shanghai 200438, China}
\affiliation{Shanghai Qi Zhi Institute, Shanghai 200030, China}
\author{Jue Nan$^{\dagger}$} 
\affiliation{State Key Laboratory of Surface Physics, Institute of Nanoelectronics and Quantum Computing, and Department of Physics, Fudan University, Shanghai 200438, China}
\affiliation{Shanghai Qi Zhi Institute, Shanghai 200030, China}
\author{Tao Zhang}
\affiliation{Department of Physics and State Key Laboratory of Low Dimensional Quantum Physics, Tsinghua University, Beijing, 100084, China}
\author{Xingze Qiu}
\affiliation{State Key Laboratory of Surface Physics, Institute of Nanoelectronics and Quantum Computing, and Department of Physics, Fudan University, Shanghai 200438, China}
\affiliation{Shanghai Qi Zhi Institute, Shanghai 200030, China}
\author{Wenlan Chen}
\email{cwlaser@ultracold.cn}
\affiliation{Department of Physics and State Key Laboratory of Low Dimensional Quantum Physics, Tsinghua University, Beijing, 100084, China}
\affiliation{Frontier Science Center for Quantum Information, Beijing, 100084, China}
\author{Xiaopeng Li}  
\email{xiaopeng\underline{ }li@fudan.edu.cn}
\affiliation{State Key Laboratory of Surface Physics, Institute of Nanoelectronics and Quantum Computing, and Department of Physics, Fudan University, Shanghai 200438, China}
\affiliation{Shanghai Qi Zhi Institute, Shanghai 200030, China}
\affiliation{Shanghai Research Center for Quantum Sciences, Shanghai 201315, China}

\begin{abstract}
Computing finite temperature properties of a quantum many-body system is key to describing a broad range of correlated quantum many-body physics from quantum chemistry and condensed matter to thermal quantum field theories. Quantum  computing with rapid developments in recent years has a huge potential to impact the computation of quantum thermodynamics. 
To fulfill the potential impacts, it is crucial to design quantum algorithms that utilize the computation power of the quantum computing devices. 
Here we present a quantum kernel function expansion (QKFE) algorithm for solving thermodynamic properties of quantum many-body systems. In this quantum algorithm, the many-body density of states is approximated by a kernel-Fourier expansion, whose expansion moments are obtained by random state sampling and  quantum interferometric measurements. As compared to its classical counterpart, namely the kernel polynomial method (KPM), QKFE has an exponential advantage in the cost of both time and memory. In computing low temperature properties, QKFE becomes inefficient, as similar to classical KPM. To resolve this difficulty, we further construct a thermal ensemble iteration (THEI) protocol, which starts from the trivial limit of infinite temperature ensemble and approaches the low temperature regime step-by-step. 
For quantum Hamiltonians, whose ground states are preparable with polynomial quantum circuits, THEI has an overall polynomial complexity. We demonstrate its efficiency with applications to one and two-dimensional quantum spin models, and a fermionic lattice. With our analysis on the realization with digital and analogue quantum devices, we expect the quantum algorithm is accessible to current quantum technology.

\end{abstract}

\maketitle

\section{Introduction}

The computation of thermodynamic quantities of quantum Hamiltonians is at the core of simulating correlated electrons in quantum materials and complex molecules~\cite{2020_Bauer_Review,2020_McArdle_RMP}.  The exponential complexity in treating a large number of entangled degrees of freedom on a classical computer prevents accurate determination of macroscopic physics~\cite{1929_Dirac,Feynman}, causing a generic challenge to our quantitative description of a broad range of strongly correlated quantum phases from quantum magnetism~\cite{2008_Sachdev_NatPhys}  and 
high T$_{\rm c}$ superconductivity~\cite{2015_Keimer_Nature} to neutron star matters~\cite{2020_Mann_Nature}.

With controllable  quantum systems, one way that has been carried out is to synthesize analog Hamiltonian models and  extract thermodynamic properties by preparing experimental systems at thermal equilibrium~\cite{2004_Cirac_Zoller,2008_Stringari_RMP,2014_Nori_RMP}. Strongly correlated physics such as Mott-superfluid transition~\cite{1989_Fisher_PRB,2002_Bloch_Nature}, unitary Fermi gas~\cite{2004_Jin_PRL}, and antiferromagnetism~\cite{2015_Hulet_AFM,2017_Bloch_Science,2017_Greiner_Nature} have been accomplished with cold atoms.  With rapid advancement of programmable quantum devices in the last several years such as superconducting qubits~\cite{2019_Google_Nature,2021_Google_Science,2021_Zhu_Science,2021_Xiaobo_PRL}, trapped ions~\cite{2017_Monre_Nature53,2019_Zoller_Nature}, entangled photons~\cite{2013_Walther_NatPhoton,2013_Crespi_NatPhoton,2013_White_Science,2013_Spring_Science,2020_Pan_Science} and Rydberg atoms~\cite{2016_Browaeys_Nature,2017_Lukin_Nature,2019_Browaeys_Science,2021_Lukin_Science}, there have been growing research interests in developing algorithmic approaches for quantum simulations~\cite{Temme2011,Yung754,Motta2020,2020_Cohn_PRA,PRXCirac,2021_Shtanko_arXiv}. 
Much progress has been made for determining ground states considering variants of quantum phase estimation~\cite{2005_AspuruGuzik_Science,2019_Cirac_JMP}, adiabatic Hamiltonian evolution~\cite{2001_Farhi_Science,2008_Aharonov_AQC} and variational quantum circuits~\cite{2014_Peruzzo_NC,2021_Cerezo_NRP,2020_Bauer_Review,2020_McArdle_RMP}. 
Quantum algorithms for finite temperature quantum simulations have also been proposed using generalized Metropolis sampling~\cite{2009_Wocjan_PRL,Temme2011,2017_Chowdhury_Quantum,PRXCirac}, quantum Lanczos methods~\cite{Motta2020}, 
and variational thermofield double state algorithms~\cite{PRLWu19,ArXGV19,Hsieh2020PNAS}. 
However, until now, finite temperature quantum simulation algorithms are relatively scarce as compared to the ground state computation. Efficient computation methods for free energy and thermal entropy, which are crucial for determining thermodynamics, are particularly lacking and in great demand.

In this work, we introduce a quantum kernel function expansion (QKFE) algorithm where the energy dependence of observables and many-body density-of-states (DOS) are represented by Fourier series. We show the expansion moments can be measured by quantum circuits with polynomial cost, achieving an exponential quantum  advantage over the classical analogue, namely, the kernel polynomial method (KPM)~\cite{KPM_RMP}. 
{ 
The QKFE quantum circuit is fully deterministic, i.e., free of variational optimization, in sharp contrast to variational quantum algorithms~\cite{PRLWu19,ArXGV19,Hsieh2020PNAS} and quantum Lanczos methods~\cite{Motta2020,2019_Yuan_npjQI,2020_Siopsis_npjQI,2022_Zeng_npjQI}.  
The infamous barren plateau problem~\cite{2018_McClean_NC,2021_Wiebe_PRXQ} is thus completely absent with QKFE. 
} 
The overall complexity of QKFE is exponential in approaching low-temperature properties of a generic Hamiltonian, which is a corollary of Hamiltonian QMA completeness~\cite{2006_Kempe_Complexity,2014_Cubitt_Complexity,2015_Shin_Complexity}. We further develop a thermal ensemble iteration protocol based on  QKFE with Hamiltonian evolutions, which computes thermodynamic quantities such as local observables, free energy, and thermal entropy with polynomial complexity, provided that the ground state of the Hamiltonian can be prepared at a polynomial cost.

\begin{figure*}[htp]
    \centering
    \includegraphics[scale=0.65]{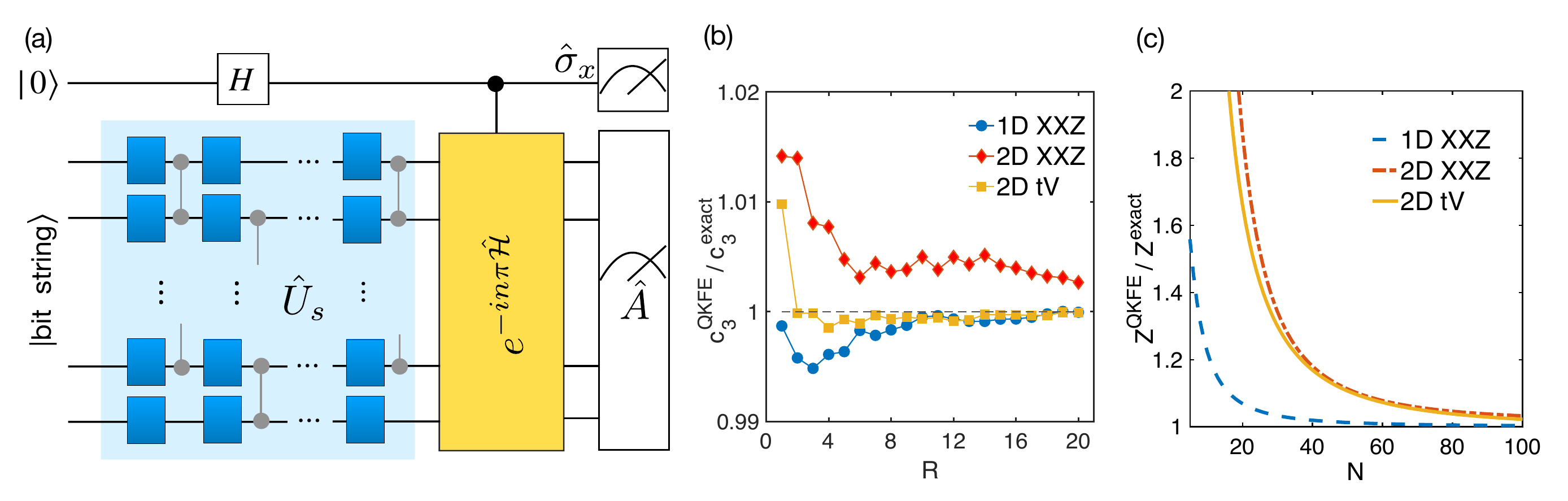}
    \caption{The quantum kernel function expansion algorithm. (a), the illustration of quantum circuits for measuring Fourier expansion moments. 
    (b), the convergence of one typical expansion moment, $c_3$, with increasing the number of random states, $R$.  
    (c), the convergence of partition function with increasing the expansion cutoff, $N$.
    {The blue, red and orange lines correspond to the results of 1D-XXZ (Eq.~\eqref{eq:1DXXZ}), the 2D-XXZ (Eq.~\eqref{eq:2DXXZ}), and the t-V (Eq.~\eqref{eq:tV}) models, respectively.}
    In (b,c), the QKFE results of the expansion moment $c_3$ and the partition function $Z$, ($c_3^{\rm QKFE}$ and $Z^{\rm QKFE}$) are normalized by their exact values. 
    Here we choose $L = 18$ for the 1D model, and a square lattice with $4\times4$ geometry for the 2D models. The temperature is fixed at $T=3$.}
    \label{fig:f1}
\end{figure*}

\section{Quantum Kernel Function Expansion}
\label{sec:QKFE} 

Our QKFE algorithm has been inspired by the classical KPM.  
Considering a many-body system with Hamiltonian $\hat{H}$, a physical quantity that is natural for KPM to compute is DOS~\cite{KPM_RMP}, which is defined as $\rho (E) =\frac{1}{D}\sum_{i} \delta(E-E_i)$, with $E_i$ the eigenvalues, $D$ the Hilbert space dimension. 
 For convenience in theoretical treatment, the energy spectra are assumed to be bounded between $E_{\rm min}$ and $E_{\rm max}$. A dimensionless energy 
  \be 
 \epsilon \equiv (E-E_{\rm min})/(E_{\rm w} + 0^+)\in (0, 1)
 \label{eq:dimlessE} 
 \ee 
 is introduced accordingly with $E_{\rm w}$ being $E_{\rm max}-E_{\rm min}$. We then have a rescaled Hamiltonian, $\hat{ {\cal H}} = (\hat{H} -E_{\rm min}\mathbb{1} ) /(E_{\rm w}+0^+) $. 
 This rescaling can always be performed for a lattice Hamiltonian having a finite Hilbert space dimension. 
 In classical KPM, the DOS is expressed in terms of a Chebyshev polynomial expansion, whose expansion moments can be computed at a cost linear to  the Hilbert space dimension.  
 This approach has been used in classical computing for finite energy properties of large matrices, and has  accomplished  a great success in solving non-interacting Anderson localization problems~\cite{KPM_RMP}.  
 However, its cost in time and memory both scales exponentially with the number of degrees of freedom---the Hilbert space dimension is  $D=2^L$ for a system of $L$ qubits, which has limited its application in simulating more complex quantum many-body systems.

In this section, we present an efficient quantum algorithm for computing the expansion moments, that has an exponential quantum speedup over the classical KPM.

\subsection{QKFE algorithm}
Instead of Chebyshev polynomial expansion as used in the classical KPM, in our QKFE algorithm we perform a Fourier expansion for the DOS, 
\be
\rho(\epsilon) = c_0 + 2 \sum_{n=1}^{N-1} c_n \cos (n\pi \epsilon), 
\label{eq:rhoexpansion} 
\ee 
for the Fourier moments are more convenient to fit into quantum computing than the Chebyshev polynomial expansion. 
We have introduced a large-moment cutoff $N$ for the expansion. 
By writing the Fourier moments in the form of, 
\be 
c_n = { \frac{1}{D} } {\rm Re} \left\{ {\rm Tr} \left[ e^{-in\pi \hat{ {\cal H}} } \right] \right\}, 
\label{eq:moment} 
\ee 
we find these moments can be obtained efficiently by  a quantum circuit shown in Fig.~\ref{fig:f1}(a), 
{which contains $L$ number of physical qubits and one  ancilla qubit.}  
The step of averaging ${\rm Tr} [\ldots] /D$ is performed by sampling Haar random states, whose computation efficiency relies on quantum typicality~\cite{Popescu2006,Sheldon06,Jochen09}. Despite the difficulty of preparing exact Haar randomness, it can be approximated by relatively shallow circuits~\cite{Boixo2018,Joseph03,Oliveira07,Richter21,2022_Zoller_SFFchaos}.

The procedure for measuring $c_n$ involves three steps. The first step is to choose $R$ number of random product states as the circuit input and scramble these states by performing {local random unitary operations} $\hat{U}_s$. The second step is to apply a  control unitary  operation, 
\be 
|0\rangle \langle 0|\otimes I+|1\rangle \langle 1| \otimes e^{-in\pi \hat{ {\cal H}}}
\label{eq:controlU} 
\ee 
across the ancilla qubit and the system. 
We then take measurements. 
The measurement outcomes of $\hat{\sigma}_x$ on the ancilla qubit average to the $c_n$ moment. 

Besides DOS, the energy dependence of local observables can also be obtained efficiently using the same quantum circuit (Fig.~\ref{fig:f1}(a)). We consider a general local observable, $\hat{A}$, which for example could represent spin polarization or  correlation functions. Its energy dependence is given by 
$\alpha (\epsilon) = \langle \epsilon | \hat{A} | \epsilon\rangle$, with $|\epsilon\rangle$ an eigenstate of $\hat{\cal H}$ with energy $\epsilon$. 
The Fourier expansion reads as 
\be 
\alpha (\epsilon) = d_0 +  2 \sum_{n=1}^{N-1} d_n \cos (n\pi \epsilon), 
\label{eq:Aexpansion} 
\ee 
with the moments 
\be 
d_n = {\frac{1}{D} } {\rm Re} \left\{ {\rm Tr} \left[ \hat{A} e^{-in\pi \hat{{\cal H}} }  \right]  \right\}. 
\label{eq:dmoment} 
\ee 
This implies the $d_n$ moments can be measured by the same quantum circuit as $c_n$. 
{The measurement outcome of the tensor product of  $\hat{\sigma}_x$ (ancilla) and  $\hat{A}$ (local observables) at the final state of quantum circuit averages to the $d_n$ moments.}

Having the Fourier moments $c_n$ and $d_n$ computed by the quantum circuit, we reconstruct the functions $\rho (\epsilon)$ and $\alpha(\epsilon)$. 
With the energy dependence of DOS  and local observables  computed, 
the partition function $Z(\beta) = {\rm Tr} \left[ e^{-\beta \hat{H}} \right]$ as a function of inverse temperature $\beta$, and the canonical ensemble average 
$
A(\beta) = {{\rm Tr} \left[ \hat{A}e^{-\beta \hat{H} }\right]} /Z(\beta), 
$
are then  given by 
\bea 
Z(\beta) &=&   {\int_{0}^{1} e^{-\beta E_{\rm w} \epsilon }\rho (\epsilon) d\epsilon},   \\ 
 A(\beta) &=&   \frac{1} {Z(\beta)} 
		{\int_{0}^{1} e^{-\beta E_{\rm w} \epsilon }{\alpha(\epsilon) }d\epsilon}.  
\eea
With the partition function, all thermodynamic quantities such as free energy and thermal entropy can then be obtained~\cite{pathria}.

In the physical implementation of our QKFE algorithm, the quantum circuit in Fig.~\ref{fig:f1}(a) can be further decomposed into local quantum gates by considering Trotterization, for which the circuit depth in extracting the Fourier moments $c_n$ and $d_n$ scales as {\bf $O (n\delta_t ^{-1})$ }, with $\delta_t$ the Trotterization step. One specific example is provided later in Section~\ref{sec:trotter}. 
In comparing with the classical analogue, namely the classical KPM, our QKFE algorithm has an exponential speedup in computing the expansion moments---the time cost for classical KPM is exponential, whereas
it is polynomial in QKFE. 
{
We emphasize here that the QKFE algorithm is free of variational optimization as its quantum circuit is  fully deterministic. This makes QKFE algorithm rather unique in comparison with variational quantum algorithms~\cite{PRLWu19,ArXGV19} and quantum Lanczos methods~\cite{Motta2020,2019_Yuan_npjQI,2020_Siopsis_npjQI,2022_Zeng_npjQI}, 
for which the variational optimization could be costly in practical computation, and sometimes encounters the infamous barren plateau problem~\cite{2018_McClean_NC,2021_Wiebe_PRXQ}. 
}

\begin{figure*}[htp]
    \centering
    \includegraphics[scale=0.52]{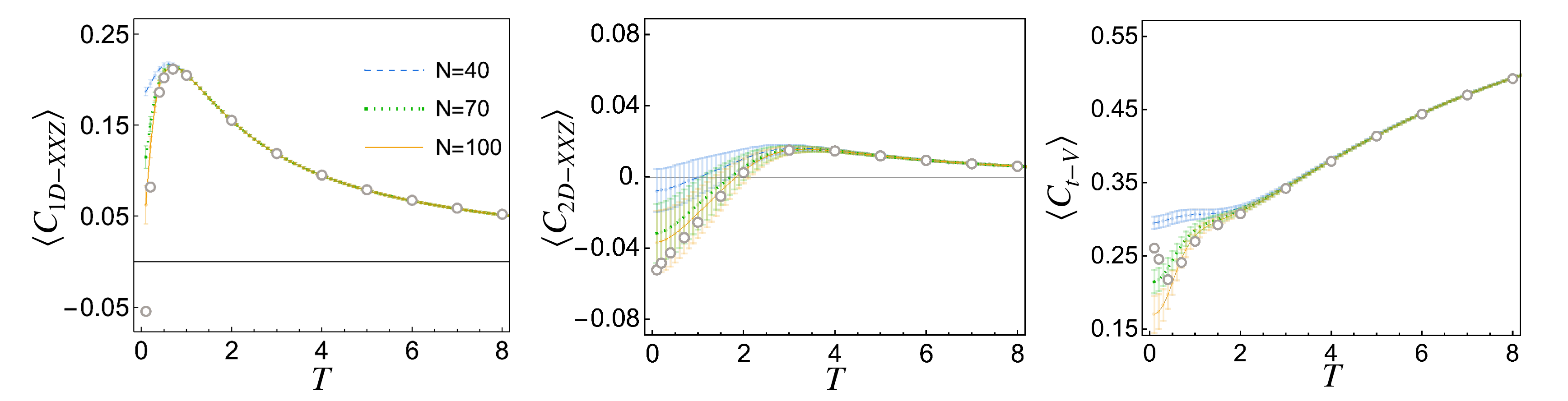}
    \caption{Finite temperature correlations with the QKFE algorithm.  The three panels from left to right correspond to 1D-XXZ, 2D-XXZ, and t-V models. For the 1D-XXZ and t-V models, we choose $R=20$ in sampling random states, and for the 2D-XXZ model, we choose $R=400$. The lines represent the numerical results by QKFE. The colored shadows surrounding  these lines are sampling errors. The circles show the exact values for comparison. }
    \label{fig:f2}
\end{figure*}

\subsection{Uniform convergence by kernel function expansion} 
\label{sec:uniform}

In reconstructing the functions $\rho (\epsilon)$ and $\alpha (\epsilon)$, we need to correct the moments by multiplying the Jackson kernel in order to damp out  the cutoff induced Gibbs oscillations. 
It is well-known that to approximate an analytic function $F(\epsilon)$, 
the $N$-th order Fourier series expansion
$
 F_{N}(\epsilon)=c_0+2\sum_n ^{N-1} c_n  \cos(n\pi\epsilon) 
$ 
has a norm convergence. However, uniform convergence is required here for computing the energy dependence of local observables and DOS. 
We apply kernel functions to the Fourier expansion. 
For a continuous function ${f} (x)$ with $x\in(-1,1)$, it has been shown in classical KPM analysis~\cite{KPM_RMP} that the  kernel-function corrected $N$-th order Chebyshev expansion 
$ 
f_N(x) = \tilde{c}_0+ 2\sum_{n=1} ^{N-1} \tilde{c}_n T_n (x) 
$ 
has a uniform convergence to $f(x)$, 
with 
\bea 
\textstyle \tilde{c}_n &=& \textstyle h_n \int_{-1} ^{1} \frac{f(x) T_n (x) }{\pi \sqrt{1-x^2} }  d x,  \nn  \\ 
 \textstyle h_{n}&=& \textstyle\frac{1}{N+1} \left[ (N-n+1) \cos \frac{\pi n}{N+1}+\sin \frac{\pi n}{N+1} \cot \frac{\pi}{N+1} \right]
 \,\,\, \nn \\ 
 \textstyle T_n  &=& \textstyle \cos \left[ n\, {\rm arccos} (x) \right],  \nn 
\eea 
where $h_n$ is the Jackson kernel~\cite{Jackson1912}. Approximating $f(x)$ by $f_N(x)$ has an error~\cite{KPM_RMP}, 
$
||f(x)-f_N(x)||_\infty \sim w_f (1/N), 
$ 
with $w_f (\delta) ={\rm max}|f(x)-f(y)|_{|x-y| \le \delta}$. 
This can be interpreted as an error at the order of  $O(1/N)$.

The Fourier expansion used in our work is related to the Chebyshev expansion by taking $x = \cos (\pi \epsilon)$, and $F(\epsilon) = f(\cos(\pi\epsilon))$.  For $\epsilon \in (0, 1)$, we have $T_n(x) = \cos (n\pi \epsilon)$. 
It follows immediately that 
\be 
||F(\epsilon) - F_N ( \epsilon) || _\infty \sim w_f (1/N),
\ee 
with 
the expansion moments $c_n$ corrected by the Jackson kernel, i.e., 
\be 
c_n\to \tilde{c}_n = c_n h_n .
\ee 
We thus conclude the kernel Fourier series expansion has uniform convergence with error $O(1/N)$.  
This correction applies the same way to the expansion of DOS and local observables.

\subsection{Measurement cost of the expansion moments} 
\label{sec:measurement}

In the QKFE algorithm, the Fourier moments $c_n$ are obtained  by 
averaging over the measurement outcome of  $\hat{\sigma}_x$ on the ancilla qubit. With $K$ times of quantum projective  measurements, the error on $c_n$ scales as $1/\sqrt{K}$. Denoting the quantum shot noise on the single-shot measurement of $c_n$ as $\eta_n$,  
the induced error on the density of states is 
\be
\mathcal{E} = 2\sum_{n=1}^{N-1} \eta_n h_n \cos (n\pi \epsilon), 
\ee
whose statistical  variance is  
\be 
{\rm Var}(\mathcal{E}) = 4\sum_{n=1}^{N-1} {\rm Var} (\eta_n)[h_n \cos (n\pi \epsilon)]^2.
\ee 
Since both of  ${\rm Var} (\eta_n)$ and $h_n$ are bounded, the variance of $\mathcal{E}$ scales as  ${\rm Var}(\mathcal{E})\sim N$. 
Averaging $K$ times, the measurement precision on the density of states is then $O(\sqrt{N/K})$. 
Since the truncation error in the Fourier expansion is $O(1/N)$ as discussed above, it is reasonable to demand the same scaling on the measurement precision, which then implies a requirement on the number of repeated projective  measurements 
\be 
K\sim N^3.
\ee  

In calculating other  observables, the requirement on the measurement cost is the same as the density of states, according to the  expansion in Eq.~\eqref{eq:Aexpansion}.

\subsection{Numerical demonstration on spin and fermion models} 
\label{sec:QKFENumeric} 

To benchmark the overall performance of our QKFE, we apply this algorithm to three lattice models including a one-dimensional (1D) spin-1/2 XXZ chain, 
\begin{equation}
\label{eq:1DXXZ} 
 \hat{H}_{1D-XXZ}= \frac{1}{2}\sum_j \hat{ \sigma}_{j}^{x}\hat{\sigma}_{j+1}^{x}+\hat{\sigma}_{j}^{y}\hat{\sigma}_{j+1}^{y}+\Delta\hat{\sigma}_{j}^{z}\hat{\sigma}_{j+1}^{z},
\end{equation}
a  two-dimensional (2D)  XXZ model, 
\begin{equation}
\label{eq:2DXXZ} 
\hat{H}_{2D-XXZ}=\sum_{<i,j>}\hat{\sigma}_{i}^{x} \hat{\sigma}_{j}^{x}+ \hat{\sigma}_{i}^{y}\hat{\sigma}_{j}^{y}+\Delta' \hat{ \sigma}_{i}^{z}\hat{\sigma}_{j}^{z},
\end{equation}
and a  2D t-V model of spinless fermions,        
\begin{equation}
\label{eq:tV} 
 \hat{H}_{tV}=- \sum_{<i,j>} \hat{c}_{i}^{\dagger} \hat{c}_{j}+\hat{c}_{j}^{\dagger}\hat{c}_{i}+V\sum_{<i,j>}\hat{n}_{i}\hat{n}_{j},
\end{equation}
with the open boundary condition adopted. 
We emphasize that our QKFE algorithm is generic in performing finite temperature quantum simulations---it is not restricted to solving these three models. We deliberately choose both spin and fermion models here for benchmarking in order to confirm QKFE indeed applies generically to different quantum Hamiltonian systems.

In the numerical tests, we choose $\Delta = -0.9$, $\Delta' = -0.5$, and $V=2$. 
The convergence is observed at $R \to 20$, and $N \to 100$.
For local observables, we examine 
 $C_{1D-XXZ}\equiv \hat{\sigma}_{1}^{z} \hat{\sigma}_{2}^{z}$, 
 $C_{2D-XXZ}\equiv \hat{\sigma}_{11}^{z} \hat{\sigma}_{22}^{z}$ and 
 $C_{tV}\equiv \hat{ n}_{11} \hat{n}_{22}+\hat{n}_{11} \hat{n}_{33}+\hat{n}_{11}\hat{n}_{44}$, 
 for the 1D-XXZ, 2D-XXZ, and t-V models, respectively. 
We have also checked  other observables and find similar behavior as presented here.  
Fig.~\ref{fig:f2} shows the performance of QKFE in a broad temperature range. 
It is apparent that the quantum algorithm performs well in the high temperature regime for all three models. 
In the low temperature regime, QKFE is no longer reliable, producing substantial computation errors.  
The large sampling error implies a large number of $R$ is required at low temperature. The sizable discrepancy between the QKFE  and exact calculation indicates a larger cutoff $N$ is also  needed to approximate the functions in Eqs.~(\ref{eq:moment}, \ref{eq:dmoment}) at low temperature.

The inefficiency of QKFE at low temperature can be attributed to two aspects. Firstly, the low energy states of the many-body Hamiltonian only make an exponentially small contribution to the expansion moments. Consequently, it is inevitable to sample an exponential number ($R$) of random states because an exponential precision would be required for the moments. Secondly, the {DOS} at low energy is exponentially smaller as compared to high energy. This makes it difficult for the Fourier expansion to approximate the entire energy window. 
The exponential cost at low energy regime has also been observed in other finite temperature algorithms evaluating partition  functions~\cite{2020_block_encoding,2021_one_clean_qubit}. 
Nonetheless, the exponential speedup in QKFE in computing the moments over the classical KPM remains valid, because these two aspects are also present for classical KPM in simulating low-energy many-body physics.   
In fact, the exponential time cost for generic low-temperature  quantum simulations is a corollary of  Hamiltonian QMA completeness~\cite{2009_Wocjan_PRL,Temme2011,2017_Chowdhury_Quantum}.

\section{Thermal Ensemble Iteration}
We further develop a polynomial quantum algorithm for a restricted class of Hamiltonian assuming that its ground state determination belongs to BQP~\cite{2014_Cubitt_Complexity}.  
An efficient scheme is provided for preparation of an excited state with finite energy density. 
We construct a thermal ensemble iteration protocol, and show that the thermodynamic quantities such as free energy and thermal entropy can be obtained with polynomial cost  by acting our  QKFE algorithm iteratively on the finite-energy quantum states. We emphasize that the THEI protocol provides a quantum algorithm for generic quantum Hamiltonian models even for those not belonging to BQP. It is an efficient quantum algorithm for Hamiltonian models whose ground states are preparable with quantum circuits at polynomial cost.

\subsection{Preparation of finite energy quantum states} 
\label{sec:preparefiniteE}
For a Hamiltonian  in BQP ($\hat{H}_{\rm BQP}$), it is guaranteed that the ground state can be reached by 
{polynomial-depth quantum circuits} that involves one- and two-qubit gates~\cite{2014_Cubitt_Complexity}. 
We choose a random product state as the initial state of this quantum circuit, whose energy is typically matching the infinite temperature ensemble, i.e., $E(\beta = 0)$. 
Here,  $E(\beta)$ is the thermal ensemble average with respect to $\hat{H}_{\rm BQP}$. 
This type of product state can be efficiently achieved due to the exponential dominance of infinite-temperature states in the quantum many-body Hilbert space.
The output of the quantum circuit is the ground state of $\hat{H}_{\rm BQP}$ with energy $E(\beta = \infty)$. 
We split the quantum circuit into multiple steps with each step containing one single-qubit or two-qubit gate only. The gate at $p$-th step is denoted as  $ \hat{U}_p$, and the quantum state at this step is $|\alpha(p)\rangle$. 
The energy at the $p$-th step is given by $E_p = \langle \alpha(p) |\hat{H}_{\rm BQP}| \alpha(p) \rangle$. 
By physical intuition, it is reasonable to assume the energy disturbance produced at each step, for instance $|E_{p}-E_{p-1}|$ from the step-$(p-1)$ to step-$p$, is upper-bounded by a constant $\Delta E_{\rm ub} $ independent of the system size. The energy density difference between two successive steps is then infinitesimal [$O(1/L)$] in the thermodynamic limit. This implies that a quantum state with intermediate energy (inbetween the ground state and infinite temperature ensemble average) can be prepared by choosing a proper intermediate $p$-step in the polynomial-depth quantum circuit preparing the ground state. The energy density resolution of this scheme for preparing an excited state with a given energy is $\Delta E_{\rm ub}/L$.

Now, we show the energy disturbance caused by one step of quantum gate operation indeed has a rigorous upper bound for a k-local Hamiltonian. The Hamiltonian we consider has a general form of $\hat{H}_{\rm BQP} = \sum_{l} \hat{H}_l$, which acts on  $L$ qubits. 
Without loss of generality, we assume $H_l$ is a Pauli operator of the form  $\hat{H}_l = J \hat{\sigma}_{i_1} \cdots \hat{\sigma}_{i_{|h_l|}}$, 
with $h_l$ representing the set of qubits that $\hat{H}_l$ acts on. 
To proceed, we expand the quantum state $|\alpha (p-1)\rangle$ at the step-($p-1$) in the computation basis $|{\bf z} \rangle$, 
$| \alpha(p-1)\rangle = \sum_{\bf z}  \psi_{\bf z}  |{\bf z}\rangle$.  It follows that 
$| \alpha(p)\rangle = \sum_{\bf z}  \psi_{\bf z} |\tilde{\bf z}\rangle$,  
with $|\tilde{\bf z}\rangle \equiv \hat{U}_p |{\bf z} \rangle$. 
Since $\hat{U}_p$ represents a one- or two-qubit gate, the state $|\tilde{\bf z}\rangle$ is different from $|{\bf z} \rangle$ only within a local region, $g_p$, defined to be the set of qubits that $\hat{U}_p$ acts on.  
The energy difference $\Delta E_p = E_p - E_{p-1}$ has a form of 
\bea
\Delta E_p =\sum_l \sum_{ {\bf z}_1, {\bf z}_2}  { \psi_{{\bf z}_2} ^* \psi_{{\bf z}_1}  
\left[ \langle \tilde{{\bf z}}_2 |\hat{H}_l| \tilde{ {\bf z}}_1  \rangle-\langle {\bf z}_2 | \hat{H}_l| {\bf z}_1 \rangle \right]. }   \nonumber
\eea
The difference in the summation is finite only when $h_l \cap g_p  \neq \emptyset$. 
We then have 
\bea
\label{eq:whole}
|\Delta E_p| 
\le \sum_{\substack{
        l \\ (h_l \cap g_p  \neq \emptyset)}}  \sum_{{\bf z}_1}  \sum_{\substack{ {\bf z}_2  \in \Xi_{l,{\bf z}_1} }} |\psi_{{\bf z}_2}| |\psi_{{\bf z}_1} |\nonumber \\
         \times \left[ \left| \langle \tilde{\bf z}_2 |\hat{H}_l| \tilde{\bf z}_1  \rangle-\langle {\bf z}_2 |\hat{H}_l| {\bf z}_1 \rangle \right| \right]. \nn 
\eea
Here, the set $\Xi_{l, {\bf z}_1}$ contains all ${\bf z}_2$ configurations  that only differ from  ${\bf z}_1$ in the local region $h_l$.  
In the following, the constrained summation over $l$, ${\bf z}_1$, and ${\bf z}_2$ as restricted by  $h_l \cap g_p  \neq \emptyset$ and ${\bf z}_2 \in \Xi_{l, {\bf z}_1}$, will be denoted as 
$\sum'_{l,{\bf z}_1, {\bf z}_2} $ to save writing. Summing over $l$ with the restriction $h_l \cap g_p  \neq \emptyset$ is denoted as $\sum_l '$ correspondingly. 
Since the quantity in the bracket is bounded by 
$
\left| \langle \tilde{{\bf z}}_2 |\hat{H}_l| \tilde{\bf z}_1 \rangle-\langle {\bf z}_2 |\hat{H}_l| {\bf z}_1 \rangle \right| \le 2|J|,
$
we have 
\[
|\Delta E_p| \le 2|J| \sum_{l, {\bf z}_1, {\bf z}_2}   ' |\psi_{{\bf z}_1} | |\psi_{{\bf z}_2} |. 
\]
Using the inequality 
\bea
\sum'_{l,{\bf z}_1, {\bf z}_2}}  {  |\psi_{{\bf z}_1} | |\psi_{{\bf z}_2} | 
\le \frac{1}{2}\sum_{l,{\bf z}_1, {\bf z}_2}'  |\psi_{{\bf z}_1} |^2 + |\psi_{{\bf z}_2} |^2  
= \sum_l ' 2^{|h_l|}, \nonumber
\eea
we obtain 
\be
|\Delta E_p| \le \Delta E_{\rm ub} = \gamma |J| 
\ee 
with the constant 
$
\gamma =
 \sum_l' 2^{|h_l|+1} . 
$
The energy disturbance $\Delta E_p$ has an upper bound that is independent of the system size $L$ for a k-local Hamiltonian.

We remark here that the above analysis applies generically for k-local Hamiltonians,  including those not belonging to BQP. This means the preparation scheme described here for finite energy quantum states is applicable to all k-local Hamiltonians. 
Its polynomial complexity relies on assuming the ground state is preparable at polynomial cost. Our finite-energy-state preparation scheme implies that finite-temperature quantum simulation is in general at most as complicated as the ground state quantum simulation.

\begin{figure*}[htb]
    \centering
    \includegraphics[scale=.95]{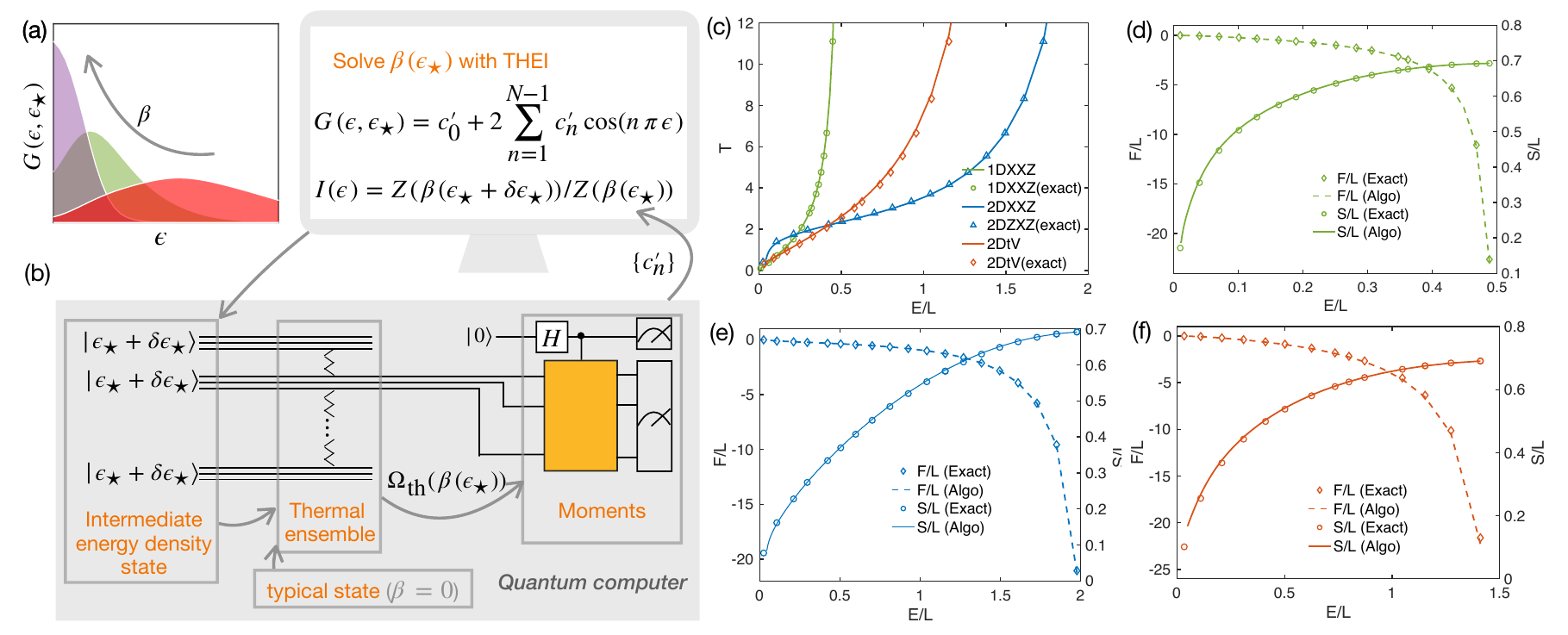}
    \caption{ Free energy and thermal entropy with THEI protocol. 
     (a,b), schematic illustration of the THEI protocol.  (c,d,e,f), the computed thermodynamic quantities by the THEI protocol. We study 1D-XXZ, 2D-XXZ, and 2D t-V fermion models, to demonstrate the generic applicability of our quantum algorithm. 
    (c) shows the energy density dependence of inverse temperature, $\beta (E)$. 
    (d,e,f), the free energy and thermal entropy as a function of energy density, corresponding to 1D-XXZ, 2D-XXZ, and t-V models, respectively. 
    The lines represent the results of THEI and the symbols, $\circ$, $\bDiamond$, $\triangle$, show the exact values for comparison. 
    Here we choose an expansion order $N = 100$. The difference  between the THEI and exact results is barely noticeable, and can be further improved by using a larger expansion order.}
    \label{fig:f3}
\end{figure*}

\subsection{The procedure of thermal ensemble iteration} 
With the finite energy quantum states, although the correlation functions and local observables can be measured directly, it is still a challenge to determine the thermodynamic quantities such as temperature, thermal entropy, and free energy. This challenge arises broadly for finite-temperature quantum simulations~\cite{PRXCirac,Motta2020,Temme2011}.  Here, we construct a thermal ensemble iteration protocol and show those thermodynamic quantities can be efficiently computed by an iterative running of QKFE algorithms. 

Our  THEI protocol is based on measuring the Fourier moments with respect to the canonical ensemble average, 
\be
c_n '(\epsilon_\star)  \equiv {\rm Re} \left\{  {\rm Tr} \left[\hat{ \Omega} _{\rm th} e^{-i n\pi \hat{ {\cal H}}_{\rm BQP}} \right] \right\},
\label{eq:THEI moments}
\ee 
which can be obtained by performing QKFE on a canonical ensemble $\hat{ \Omega} _{\rm th}  = e^{-\beta \hat{H}_{\rm BQP}}/Z(\beta)$ (Fig.~\ref{fig:f3}). 
Here, $\epsilon_\star$ is  ensemble average energy [rescaled according to Eq.~\eqref{eq:dimlessE}], which can be measured directly on the quantum circuit.  Since the average of unitary operator $e^{-i n\pi \hat{ {\cal H}}_{\rm BQP}}$  is now performed over the canonical ensemble instead of random states as in Sec.~\ref{sec:QKFE}, the Fourier moments $c_n'$ acquire energy ($\epsilon_\star$) dependence, or equivalently $\beta$ dependence. 
There is a one-to-one correspondence between $\epsilon_\star$ and $\beta$. This defines a function relation $\beta (\epsilon_\star)$, in spite of the difficulty to infer this relation directly on the quantum circuit.

In order to prepare the canonical ensemble, we propose preparing  multiple ($M$) copies of $\hat{H}_{\rm BQP}$, which are allowed to exchange energy by weak interactions only. Each copy is prepared in an intermediate energy density state following the protocol described in Sec.~\ref{sec:preparefiniteE}. It has been proved that the reduced density matrix of each copy is typically very close to the canonical ensemble, with a trace distance that decays exponentially with $M$~\cite{Popescu2006}. We remark here that for NISQ devices~\cite{2018_Preskill_Quantum}, which unavoidably couple to the environment anyway, preparing multiple weakly-coupled copies is expected to be unnecessary.

Having the expansion moments (Eq.~\eqref{eq:THEI moments}) measured by running QKFE with canonical ensemble as its input,  
the energy distribution function $\rho (\epsilon) e^{-\beta(\epsilon_\star)   \epsilon E_{\rm w}} /Z(\beta)$ is then approximated by 
\be  
G(\epsilon, \epsilon_\star) = {c}'_{0} (\epsilon_\star) +2\sum_{n=1}^{N-1} {c}'_{n} (\epsilon_\star) \cos (n\pi \epsilon).
\label{eq:Gapp} 
\ee 
It is apparent that 
\[
\frac{ \rho (\epsilon) e^{-\beta (\epsilon_\star) \epsilon E_{\rm w} } } {\rho (\epsilon) e^{-\beta (\epsilon_\star+ \delta \epsilon_\star) \epsilon E_{\rm w} } } 
\times \frac{e^{ \beta(\epsilon_\star) \epsilon E_{\rm w} } }
                                    {e^{ \beta( \epsilon_\star + \delta {\epsilon_\star}  ) \epsilon E_{\rm w} }} 
\]
is trivially independent of $\epsilon$ by definition. 
This simple consideration implies as yet a nontrivial condition on $G(\epsilon, \epsilon_\star)$---the function 
\be 
I(\epsilon) \equiv \frac{G(\epsilon, \epsilon_\star)  }
                        {G(\epsilon, \epsilon_\star+\delta \epsilon_\star)  } 
                        \times \frac{e^{ \beta(\epsilon_\star) \epsilon E_{\rm w} } }
                                    {e^{ \beta( \epsilon_\star + \delta {\epsilon_\star}  ) \epsilon E_{\rm w} }} 
\label{eq:Iepsilon} 
\ee 
should be independent of $\epsilon$. 
Suppose $\beta (\epsilon_\star)$ is already known, then the function $I(\epsilon)$ satisfies the $\epsilon$-independent condition only if $\beta (\epsilon_\star + \delta \epsilon_\star)$ is correct, because otherwise the incorrectness would produce an artificial exponential $\epsilon$-dependence. 
We then determine the correct value for $\beta (\epsilon_\star + \delta \epsilon_\star)$ by minimizing the $\epsilon$-dependence of $I(\epsilon)$, 
\be 
\beta (\epsilon_\star + \delta \epsilon_\star) \leftarrow 
{\rm min} \left\{  1 -  \left[ \int d\epsilon I(\epsilon) \right] ^2/\int d \epsilon I^2 (\epsilon) \right\} 
\ee 
This defines an iteration from $\beta (\epsilon_\star)$ to $\beta (\epsilon_\star +\delta {\epsilon_\star}) $, and directly produces the ratio of the partition function, $Z(\beta (\epsilon_\star + \delta \epsilon_\star) )/Z(\beta (\epsilon_\star ) )$. 
How close $I(\epsilon)$ is to a constant function in the actual computation  can be used as a self-verification indicator for whether the canonical ensemble has indeed been reached.

Since the inverse temperature $\beta (\epsilon_\star)$ is known at the infinite temperature limit, the function $\beta (\epsilon_\star)$ is then obtained  by following the iteration from $\epsilon_\star$ to $\epsilon_\star+\delta \epsilon_\star$, step-by-step (Fig.~\ref{fig:f3}(a)). 
Likewise, the partition function at $\beta = 0$ is also trivially given,  $Z(\beta =0) = D$. 
The iteration by Eq.~\eqref{eq:Iepsilon} then also produces the partition function $Z(\beta)$, from which the free energy is given by $F(\beta)  = -\log Z(\beta)/\beta $,  and the thermal entropy is given by $S(\beta) = \beta \left[ E(\beta) -F(\beta) \right] $, with Boltzman constant taken as a unit.

\subsection{Stepping complexity of the THEI protocol}
In our THEI protocol, the thermodynamic properties of a canonical ensemble are obtained by stepping from the infinite temperature limit to a finite temperature. Here, we analyze the stepping complexity of the THEI protocol. 
From the uniform convergence of the kernel Fourier expansion, the energy distribution function $\rho(\epsilon) e^{-\beta \epsilon E_{\rm w} } /Z(\beta) $ is well approximated by the expansion $G(\epsilon, \epsilon_\star)$ in the energy window $[\epsilon_\star- \sigma_\epsilon, \epsilon_\star+\sigma_\epsilon]$, with  $\sigma_\epsilon$ characterizing  the energy fluctuation of the canonical ensemble (rescaled by $E_{\rm w}$ in our notation)  that scales with the system size as $1/\sqrt{L}$~\cite{pathria}.
Within this window both the absolute and relative errors are suppressed by $O(1/N)$ (see Sec.~\ref{sec:uniform}), whereas outside this energy window, the expanded function can be exponentially small, and the relative error is no longer controllable. This sets a requirement on the stepping in the THEI protocol---the step in $\epsilon_\star$, $\delta \epsilon_\star$ should scale with  $\sigma_\epsilon$, i.e., 
\be 
\delta \epsilon_\star \sim 1/\sqrt{L}. 
\ee  
Computing a thermodynamic quantity as a function of $\beta$ using THEI then involves $O(\sqrt{L})$ runs  of QKFE. 

{
In the THEI protocol, the finite energy states have to be prepared following the scheme in Sec.~\ref{sec:preparefiniteE} repeatedly. The number of repeats scales polynomially as 
${\cal O} (N^4 \sqrt{L})$, with the measurement cost taken into account (Sec.~\ref{sec:measurement}).  
In order to reduce the quantum state preparation cost, it is worth considering  further integrating quantum non-demolition measurement~\cite{2020_Zoller_SFFQND} to our scheme in determining the expansion moments of Eq.~\eqref{eq:THEI moments}, which is expected to improve the efficiency of our THEI protocol. 
This is left for future investigation.}

\subsection{Numerical demonstration}

We apply the THEI protocol to the 1D-XXZ, 2D-XXZ, and t-V models, which have been introduced in Sec.~\ref{sec:QKFENumeric}, and examine its performance from high to low temperatures. 
The results are shown in Fig.~\ref{fig:f3}. 
We confirm the computed inverse temperature, free energy and thermal entropy by the iteration protocol matches on the exact values, with errors barely noticeable.
It is evident this protocol performs well for the entire temperature range for all the three models.
The large discrepancy observed for the original QKFE in Fig.~\ref{fig:f2} at low temperature  is no longer present with the THEI protocol (Fig.~\ref{fig:f3}(c,d,e,f)). With our  numerical demonstration on two spin and one fermion models, we expect the THEI protocol with a polynomial cost on a quantum device is generically applicable for finite temperature quantum simulation  of  quantum Hamiltonian models.

\section{Experimental realization}

One key building block for our quantum algorithms including QKFE and THEI is the control-unitary as shown in Fig.~\ref{fig:f1}. In this section we discuss its physical realization, considering both  digital quantum circuit and analog quantum simulation devices, and analyze the actual implementation cost.  

\begin{figure}[htb]
    \centering
    \includegraphics[scale=.9]{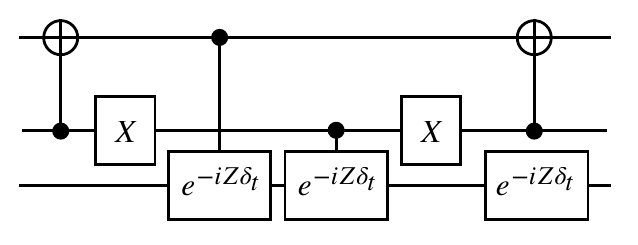}
    \caption{Quantum circuit diagram decomposition of the controlled-$ZZ$ evolution with one- and two-qubit gates.}
    \label{fig:circuit}
\end{figure}

\subsection{Realization with digital quantum circuits} 
\label{sec:trotter}
For realization of our quantum algorithms using digital quantum computation, we consider trotterizing the control unitary  in Eq.~\eqref{eq:controlU}. 
One trotter step is 
$
|0\rangle \langle 0|\otimes I+|1\rangle \langle 1| \otimes e^{-i \hat{ {\cal H}} \delta_t}, 
$
with $\delta_t$ controlling the trotter error.  
To be concrete, we consider spin models containing one- and two-body interactions. 
The control unitary corresponding to the one-body terms is realized by a control-phase gate~\cite{2021TimecrystallineEO}. 
The control unitary corresponding to the two-body terms can be constructed with three-qubit gates like C-iSWAP gates or Toffoli gates, as proposed in Ref.~\cite{2020_ciswap}.
In Fig.~\ref{fig:circuit} we provide an efficient gate decomposition using one- and two-qubit gates only for a  control unitary corresponding to the two-body terms, the controlled-$ZZ$ evolution. 
The controlled-$XX$($YY$) evolution can be decomposed in a similar way with four additional single-qubit rotation operations. 
Taking the 1D-XXZ spin chain as one example, the trotter realization of the control unitary in Eq.~\eqref{eq:controlU} involves  $15(L-1)n\pi\delta_t ^{-1}$ number of two qubit gates.  The depth of the QKFE circuit then scales as $O(\delta_t  ^{-1} N) $.
The required number of two-qubit gates would determine the size of the quantum simulation problem that can be performed on the near-term quantum devices. 
With the scheme provided here, performing QKFE on the 1D-XXZ spin chain with $L=4$, and $N = 3$, and $\delta_t = 0.2\pi $ would require a total number of $675$ two-qubit gates. 
At the same time, we expect the quantum circuit realization can be further simplified with recently developed  circuit-depth reduction techniques~\cite{Deng2020PRL,Gyongyosi2020Circuit,Fosel2021quantum}, which is left for future investigation.
An experimental demonstration of QKFE is within reasonable accessibility to superconducting qubit systems.  
For the linear scaling of  the QKFE  circuit depth in $\delta_t^{-1}$ and $N$, 
we anticipate finite temperature quantum simulation based on our proposing quantum algorithms could reach beyond classical simulation capability with near-term quantum technology.

\subsection{Realization with atom-based analog quantum simulator} 
\label{sec:RydbergExp}
Here we take the 1D-XXZ model as an example and provide an experimental realization of the control unitary in Eq.~\eqref{eq:controlU} with an atom-based quantum simulation system. We consider a system of ultracold atoms confined in a periodic optical lattice. 
The atoms are prepared in two hyperfine states with the Zeeman sublevels representing two spin-1/2 states. 
In the Mott insulator regime, the system is described by an effective XXZ Hamiltonian~\cite{2020_Ketterle_Nature}
\begin{equation}
\hat{H} = \sum_{\langle ij \rangle} [J_{xy}(\sigma_i^x \sigma_j^x+\sigma_i^y \sigma_j^y) + J_z \sigma_i^z \sigma_j^z],    
\label{eq:ATOM} 
\end{equation}
where the spin coupling are mediated by superexchange.  We have $J_{xy}=-t^2/U_{\uparrow\downarrow}$ and $J_z = t^2/U_{\uparrow\downarrow}-(t^2/U_{\uparrow\uparrow}+t^2/U_{\downarrow\downarrow})$ with $t$ the single-particle tunneling across nearby lattice sites, and $U_{\sigma \sigma'}$ the onsite Hubbard interactions.  In this system, the anisotropy $\Delta = J_z/J_{xy}$ is tunable through Feshbach resonance.

We introduce a separately controllable atom as the ancilla qubit confined  with an optical tweezer in Fig.~\ref{fig:Rydberg}. 
The ancilla qubit can be encoded by 
two hyperfine ground states dressed with Rydberg excitations. 
The control unitary corresponding to the Hamiltonian in Eq.~\eqref{eq:ATOM} is then realized by combining Raman-induced tunneling~\cite{2013_Ketterle_Harper,2013_Bloch_Harper} and Rydberg blockade~\cite{2010_Saffman_RMP,2020_Zoller_SFFQND}. 
The direct atomic tunnelings should be suppressed by adding a large enough linear tilt potential to the optical lattice. 
The Raman assisted coupling between neighboring sites is then enabled by setting the Raman detuning $\delta$ resonant with the energy offset of the nearby lattice sites. 
By using a Rydberg state as the intermediate state of the Raman transition, the Raman assisted tunneling can then be switched on and off according to the ancilla qubit.

\begin{figure}[htb]
    \centering
    \includegraphics[scale=0.9]{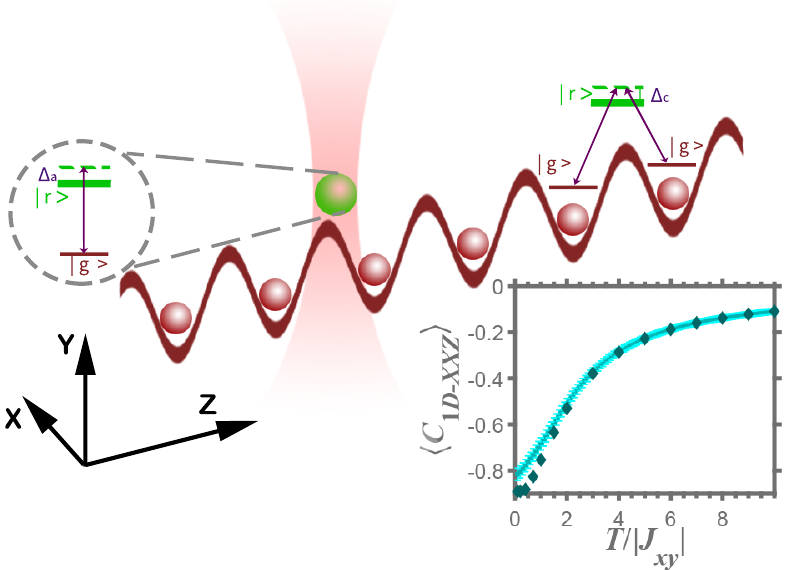}
    \caption{Experiment protocol for implementing the QKFE algorithm on analog quantum simulator based on atoms confined  in an optical lattice and an optical tweezer. This setup implements the QKFE algorithm for the 1D-XXZ model ({\it see the main text}).    Finite temperature correlations of 1D-XXZ model, $C_{1D-XXZ} = \braket{ \hat{\sigma}_1^z \hat{\sigma}_2^z} $, with QKFE (`blue line') are shown in the inset, and compared to the exact results (`diamond symbols' ). The results are calculated with $N=14$, $R=100$, using  the parameters of proposed Rydberg atom experiment in Sec.~\ref{sec:RydbergExp}.}
    \label{fig:Rydberg}
\end{figure}

To be more concrete, we consider $^{87}\mathrm{Rb}$ atoms as carriers of both the control unitary and ancilla qubits. 
The lattice depth is set as $V_x, V_y = 30E_R$ ($E_R$ as the recoil energy) in the x, y directions and $V_z = 3.6E_R$ in the z direction tilted by a spin-independent linear potential $\Delta_L = h\times 2.0\mathrm{kHz}$ per site,  with $h$ the Planck constant.  
This setup leads to a one-dimensional (1D) tilted lattice as shown in Fig.~\ref{fig:Rydberg}. 
 The Raman process is established via single-photon coupling to Rydberg state as the intermediate state~\cite{2019_JunRui_Science}, with a single-photon Rabi frequency $\Omega_c = 17 \mathrm{kHz}$ and  the single-photon detuning $\Delta_c = 300\mathrm{kHz}$. 
This set of optimized parameters lies in the Mott insulator regime and leads to isotropic superexchange interactions $J_{xy} = J_{z} = -h \times 2.78 \mathrm{Hz}$ in the absence of Feshbach resonance. 
For the ancilla qubit encoding, we use the Raman laser to selectively dress one of the two encoding hyperfine ground  states. 
The fraction of Rydberg excitations in the dressing scheme is maintained at a certain level for sufficient long lifetime. 
Here, we assume the ancilla qubit is placed properly such that the bare Rydberg interaction energy  ($ \Delta_{r-r}$) between lattice confined atoms and the ancilla qubit satisfies $\Delta_{r-r} \ge 1\mathrm{GHz}$~\cite{2005_Rydberg_interaction}.   
Then the ancilla qubit in the Rydberg dressed state would cause an energy shift $\Delta_{r-r}\Omega_a^2/(4\Delta_a^2) \ge 800\mathrm{kHz}$ on the Raman intermediate state of each atom in the 1D lattice, which consequently switches off the Raman assisted tunneling. The Raman assisted tunneling is on for the ancilla in the other encoding hyperfine state.

We choose Rydberg state $110P$ with an estimated lifetime $\tau_0 \approx 1.7\mathrm{ms}$~\cite{2009_Rydberg_lifetime,Balewski2013CouplingAS} for both Raman intermediate state and ancilla Rydberg dressing. 
The lifetime of the composite system of $L=6$ lattice qubits and the ancilla is above  $300\mathrm{ms}$ with our Rydberg dressing scheme. 
Suppose the experimental system is allowed to run for $170\mathrm{ms}$, i.e., within its lifetime, the QKFE algorithm can be performed to $N=14$ orders for the XXZ model in Eq.~\eqref{eq:ATOM} with the above superexchange interaction. 
With this setup, we calculate the finite-temperature correlation $\hat{\sigma}_1^z \hat{\sigma}_2^z$ using QKFE by averaging $R = 100$ random states and show the results in the inset of Fig.~\ref{fig:Rydberg}. 
The QKFE results agree well with the exact results having a  tiny discrepancy only in the low temperature regime with $T/|J_{xy}|<1$.

We remark here that the control unitary for 2D XXZ model can be realized in similar way using a 2D square lattice, where the linear tilt potential should be added along the diagonal direction of the lattice. The ancilla qubit can be placed near the lattice plane. 
Besides, the experimental proposal can be made even more efficient with $^{39}\mathrm{K}$ atoms~\cite{2021_Gross_RamanSidebandRydberg}, since the superexchange is naturally stronger for lighter atoms. 
In the meantime, the coupling strengths have a much  larger degree of tunability by using Feshbach resonances of $^{39}\mathrm{K}$~\cite{D_Errico_2007}.

\section{Conclusion}

We propose  a quantum kernel function expansion algorithm for finite temperature quantum simulations, where the key is to 
expand density-of-states and energy dependence of local observables by Jackson kernel corrected Fourier series. 
The QKFE algorithm is in complete absence of variational optimization, which is required in other quantum Hamiltonian algorithms such as variational quantum eigensolvers and quantum Lanczos methods. 
For a generic Hamiltonian, the QKFE algorithm has an exponential quantum advantage as compared to its classical analogue in computing expansion moments. For a BQP Hamiltonian, we equip QKFE with a THEI protocol, which constitutes an efficient finite temperature quantum simulation method for computing thermodynamic quantities such as free energy and thermal entropy with  polynomial time cost. 
For a more general Hamiltonian beyond BQP, the THEI protocol remains to be  applicable, and its time cost in performing  finite-temperature quantum simulations is comparable to finding the ground state of the Hamiltonian. In analyzing the experimental realization considering superconducting qubit and Rydberg atom quantum simulating platforms, we find the QKFE algorithm is accessible to current quantum technology.

\medskip 
\section{Acknowledgments}

We acknowledge helpful discussion with Peter Zoller, Denis Vasilyev, and Lata Kharkwal Joshi. 
This work is supported by National Program on Key Basic Research Project of China (Grant No. 2021YFA1400900), , National Natural Science Foundation of China (Grants No. 11934002, No. 92165203), Shanghai Municipal Science and Technology Major Project (Grant No. 2019SHZDZCX01), and Shanghai Science Foundation (Grants No.21QA1400500). 

$^\dagger$ These authors contributed equally to this work. 

\appendix


\bibliography{reference}
\bibliographystyle{apsrev4-1}

\end{document}